\documentclass[conference]{IEEEtran}
\usepackage{cite}
\usepackage{amsmath,amssymb,amsfonts}
\usepackage{algorithmic}
\usepackage{graphicx}
\usepackage{textcomp}
\usepackage{xcolor}
\usepackage[hyphens]{url}
\usepackage{fancyhdr}
\usepackage{tabularx,ragged2e,makecell}
\usepackage{array}
\usepackage{pifont} 
\usepackage[bookmarks=true,breaklinks=true,letterpaper=true,colorlinks,citecolor=blue,linkcolor=blue,urlcolor=blue]{hyperref}
\usepackage{xcolor, soul}
\usepackage{cleveref}
\usepackage[normalem]{ulem}
\usepackage{subcaption}
\usepackage{authblk}
\usepackage{babel}

\author{
  Akhil Shekar\\
  University of Virginia\\
  as8hu@virginia.edu\\ 
  \and
  Kevin Gaffney\\
  Microsoft\\
  kevin.gaffney@microsoft.com \\
  \and
  Martin Prammer\\
  Carnegie Mellon University\\
  mprammer@andrew.cmu.edu\\
  \and
  Khyati Kiyawat\\
  University of Virginia\\
  vyn9mp@virginia.edu\\
  \and
  Lingxi  Wu\\
  University of Virginia\\
  lw2ef@virginia.edu\\
  \and
  Helena Caminal\\
  Cornell University\\
  hc922@cornell.edu\\
  \and
  Zhenxing Fan\\
  University of Virginia\\
  fjy3ws@virginia.edu\\
  \and
  Yimin Gao\\
  University of Virginia\\
  yg9bq@virginia.edu\\
  \and
  Ashish Venkat\\
  University of Virginia\\
  venkat@virginia.edu\\
  \and
  José F. Martínez\\
  Cornell University\\
  martinez@cornell.edu\\
  \and
  Jignesh M. Patel\\
  Carnegie Mellon University\\
  jignesh@cmu.edu\\
  \and
  Kevin Skadron\\
  University of Virginia\\
  skadron@virginia.edu\\
}

\title{Membrane: Accelerating Database Analytics with Bank-Level DRAM-PIM Filtering} 

\begin{document}
\maketitle
\thispagestyle{plain}
\pagestyle{plain}


\begin{abstract}
In-memory database query processing frequently involves substantial data transfers between the CPU and memory, leading to inefficiencies due to Von Neumann bottleneck. Processing-in-Memory (PIM) architectures offer a viable solution to alleviate this bottleneck. In our study, we employ a commonly used software approach that streamlines JOIN operations into simpler selection or filtering tasks using pre-join denormalization which makes query processing workload more amenable to PIM acceleration. This research explores DRAM design landscape to evaluate how effectively these filtering tasks can be efficiently executed across DRAM hierarchy and their effect on overall application speedup. We also find that operations such as aggregates are more suitably executed on the CPU rather than PIM. Thus, we propose a cooperative query processing framework that capitalizes on both CPU and PIM strengths, where (i) the DRAM-based PIM block, with its massive parallelism, supports scan operations while (ii) CPU, with its flexible architecture, supports the rest of query execution. This allows us to utilize both PIM and CPU where appropriate and prevent dramatic changes to the overall system architecture.

With these minimal modifications, our methodology enables us to faithfully perform end-to-end performance evaluations using established analytics benchmarks such as TPCH and star-schema benchmark (SSB). Our findings show that this novel mapping approach improves performance, delivering a 5.92x/6.5x speedup compared to a traditional schema and 3.03-4.05x speedup compared to a denormalized schema with 9-17\% memory overhead, depending on the degree of partial denormalization. Further, we provide insights into query selectivity, memory overheads, and software optimizations in the context of PIM-based filtering, which better explain the behavior and performance of these systems across the benchmarks.

\nocite{*}

\end{abstract}


\section{Introduction}


Online Analytic Processing (OLAP) systems are critical technologies used to unlock the potential of  vast enterprise databases. These systems employ analytic SQL queries to transform database contents into graphs on live dashboards, generate summary reports for key performance indicators (KPIs), and trigger alerts when KPIs deviate from the norm.  In modern enterprise settings, such analytic SQL queries are also used to prepare enterprise databases for downstream machine learning (ML) pipelines (i.e., the data-heavy portions of an ML pipeline related to data cleaning and feature engineering are often done in SQL).  

Enterprise databases have consistently grown in size over the past five decades. 
Historically, Moore's Law allowed hardware performance to keep up, 
while maintaining a near-constant cost from one hardware generation to the next. 
However, it is now evident that this trajectory is no longer sustainable. Indeed, Google recently showed results from profiling its  fleet and found that BigQuery, an analytics platform, consumed about 10\% of total cycles  within the fleet, and proposed analytics as a candidate for acceleration.~\cite{google-fleet-analytics-2023-ISCA}

Furthermore, the importance of {\em in-memory} database organizations is growing rapidly for OLAP systems, including in data science and business analytics settings where complex analytic queries are often performed with a human-in-the-loop (a key driver behind the rise of DuckDB)~\cite{alliedmarketresearch}, requiring high performance on individual queries, in addition to high overall throughput. However, these workloads are often bound by the memory system's performance in conventional von Neumann-style processing systems (which dominates the server landscape on which database systems are deployed)~\cite{SirinA20}. This memory wall~\cite{memorywall} is likely to become worse over time, as memory densities are likely to grow faster than memory bus speeds (both latency and throughput impact OLAP workload performance)~\cite{olapmarket}. Even when the database does not fit in memory, smart methods of caching or staging data from disk are used by the database management system (DBMS) to keep hot data in memory. 
Thus, the CPU memory system is critical for overall query performance~\cite{seattlereport}. 


Our paper thus explores near-data processing and processing-in-mem\-ory (PIM) for analytic SQL queries. Notably absent from the prior efforts in this domain is an exploration of the different options for placing the processing at different locations within the memory architecture, in light of the impact on {\em end-to-end query execution time}. We explore placing processing elements in the channel interface and the rank, bank, and subarray levels of the memory hierarchy. We find that aggressive PIM architectures are {\em not} needed, because even modest, bank-level PIM architectures are able to significantly accelerate the PIM-friendly task of filtering the database to find the desired records---enough to make the remaining, less PIM-friendly tasks (fetching the selected records and postprocessing, i.e., aggregation, sorting, etc.) the new bottleneck. Further improvements in filtering yield minimal speedup, thanks to Amdahl's Law. 

We propose PIM that is specialized for data analytics, and filtering in particular, because this represents ``low-hanging fruit'' for an initial PIM product. Because filtering is so important, and primarily involves only simple operations over numeric and dictionary-encoded columns, the implementation can achieve high utilization of the new hardware. Furthermore, our proposed architecture is very lightweight, incurring minimal changes to the DRAM and CPU architecture, and minimal area and power overhead. Our proposed design only adds an area-optimized comparison unit to each DRAM bank and a small change in how cache line fetch and interleaving interact, and does not require any other changes to data layout. Our goal is that there should be negligible impact on chip capacity. We observe a marginal 0.1\% area overhead, and we are able to avoid the need to restructure data between PIM access and regular memory read/write, and there is no impact on conventional read/write performance. This allows the PIM feature to be completely transparent to applications or application phases that do not use PIM. Furthermore, targeting PIM avoids the chicken-and-egg problem that faces many accelerators, in which there is a lack of applications and programming models to create a ready market.  With OLAP, the SQL interface allows the PIM product to be transparent to the users and have a ready market, and the analytics market is large enough that it can likely support a specialized PIM product.  Taken together, our goal is a design that can enable low-risk adoption of PIM in commodity DRAM, so that if this design is successful, it can serve as a starting point for more sophisticated and general-purpose PIM architectures.  The main impact of adding PIM is that activating all banks in parallel leads to a 4x increase in DRAM power, requiring improved power delivery and cooling. However, end-to-end energy efficiency improves by 3.4x.


In this paper, we show that end-to-end query processing does indeed benefit from PIM and present the following contributions:

(1) We concentrate on DRAM-based PIM and explore the hardware design possibilities for moving query processing closer to the data in memory. We argue that the filtering step is both the most important and also the best fit for PIM. The options we consider are rank-level processing (via a small module on the DIMM module's circuit board), two forms of bank-level PIM, and subarray-level PIM, and evaluate their impact on end-to-end performance as well as performance relative to the extra area required to implement them. We show that the bank level provides the best combination of performance and low overhead.

(2) Inspired by a prevalent database technique called WideTable \cite{widetable}, we use denormalization and dictionary encoding to replace joins with filters, improving PIM amenability. Because full denormalization incurs prohibitive space overhead (73\% for SSB and exceeding available memory for TPC-H), we propose an approach that uses static analysis of the workload to determine which columns to denormalize. Exploring the tradeoff between space overhead and performance improvement, we find that partial denormalization with PIM filtering enables 5.9x / 6.4x speedup with only 17\% / 13\% additional space for SSB / TPC-H.

(3) We explore several dimensions of the hardware and software co-design space, and we present a variety of insights on the relationships between hardware parallelism, filter selectivity, database size, software optimization, and performance.




(4) We describe the full end-to-end implementation in DuckDB, a widely-used state-of-the-art OLAP database system \cite{duckDB}, including system integration.

\section{Background} \label{sec:background}

\subsection{OLAP database systems} \label{sec:bkg:DB}

This paper focuses on accelerating database analytics, in particular online analytical processing (OLAP), a category of workload concerned with efficiently gathering insights from large datasets. To facilitate understanding, we provide a brief overview of the aspects of OLAP database systems that are most relevant to our contributions.

\subsubsection{Database organization}

OLAP databases typically contain a vast amount of historical data that has accumulated over time. This data is typically organized into one or more large, central \textit{fact} tables and several smaller \textit{dimension} tables. Fact tables store the primary entities in the database. For example, an e-commerce company may have an \texttt{orders} fact table with one record for each purchased item, including information about the price, discount, and date of the order. Dimension tables store additional information about rows in the fact tables. For example, \texttt{product} and \texttt{customer} dimension tables may contain information about the product that was ordered and the customer that placed the order. OLAP queries typically involve filtering the records of the fact tables and dimension tables, joining the filtered records together, and then grouping, aggregating, and/or sorting the joined records to produce an informative result. For example, an OLAP query could be used to answer a question of the form, ``For each product in category X, what is the maximum discount applied to an order placed by a customer living in city Y?'' Tables generally consist of integer, decimal, and string data, which are sometimes combined to form more complex data types. Frequent strings are often dictionary-encoded into integers, and comparisons involving strings use the encoded values when appropriate. Traditional tree-based and hash-based indexes are scarce in OLAP databases due to their space and maintenance overhead, especially with a variety of queries. Instead, lightweight batch-level statistics (\textit{e.g.}, minimum and maximum in DuckDB) are used to improve filter efficiency \cite{duckdbindexes}. Databases targeting analytics are typically laid out in memory in a column-store format, in which a given column (i.e., record field) is laid out consecutively, instead of the traditional row-store, in which all fields of a record are kept together.

\subsubsection{Core operators}

OLAP database systems employ a small number of logical operators that can be combined together to execute complex queries. The input of each operator is one or more logical tables, often referred to as relations in the context of relational algebra. The output is a single logical table. We emphasize the distinction between logical and physical here, noting that the results of each operator are not necessarily materialized and are often streamed to subsequent operators in a pipelined fashion. The \textit{filter} operator (also known as \textit{selection}) returns the subset of its input rows for which some Boolean expression evaluates to true. Filters are often evaluated as part of a table scan, although scans without filters do occur. The \textit{projection} operator computes an expression on its input columns (\textit{e.g.}, multiplying two columns together). The \textit{join} operator finds matching rows between two tables based on some condition. The \textit{aggregation} and \textit{group-by aggregation} operators compute aggregate values (\textit{e.g.}, sum) for one or more groups in the input. Each logical operator has one or more physical implementations that may be specialized for a particular situation. For example, joins with equality conditions are typically executed using hash joins. Operator specialization exposes a natural entry point for integrating PIM filtering into the rest of the database system stack. We propose PIM filtering as a specialization of the filter operator. We provide additional details about system integration later in the paper.

\subsubsection{Denormalization}

Given the read-mostly and append-only nature of OLAP databases, a common method to speed up query processing is to \textit{denormalize} the database schema. This technique folds information from the dimension table(s) into the fact table so that a join is no longer needed to evaluate OLAP queries. In research, it has already become a common requirement for software-based OLAP acceleration methods~\cite{widetable, quickstep, ambit, bitweaving, byteslice}. Denormalization is now prevalent in multiple commercial products as well (\textit{e.g.}, \cite{rockset,clickhouse,starRocks}).  WideTable~\cite{widetable} is a specific, widely-used style of denormalization. Although denormalization comes at the cost of increasing the database size, dictionary-based encoding can limit this overhead (to 9-17\% in our experiments) \cite{widetable,farber2012sap, krueger2011, raman2008constant, patel2013design}.

\subsubsection{Memory performance}

OLAP queries are data-intensive, involving relatively few processor cycles per byte of input data. For example, when a query asks for all customers in a given zip code, it may scan an entire table while only applying a simple comparison operation on each input record. As CPU speed and memory size have increased faster than both the memory speed and memory bus bandwidth, OLAP query evaluation in main-memory environments (the focus of this paper) is often memory-bound~\cite{SirinA20}.

\subsection{DRAM} \label{sec:bkg:DRAM}


DRAM is available in diverse configurations, including various forms of DDR (conventional dual inline memory modules, DIMM), as well as higher-performance, more expensive formats such as GDDR and HBM,  with DDR being the prevalent choice for main memory in most server systems that would be used for OLAP workloads. 
The CPU is connected to one or more memory {\em channels}, each of which is managed by a memory controller on the processor.  The DDR channel interface is 64 bits wide.  Typically, each memory channel operates independently of others, allowing multiple channels to concurrently perform read/write operations without interference. A standard CPU is capable of supporting two or more memory channels, with high-end server systems accommodating as many as eight channels per processor.

The DDR channels are subdivided into ranks, where a rank is a set of DRAM chips that operate in parallel and can each read or write 4, 8, or 16 bits in a single operation, called the burst width. In an x8 configuration, a single chip contributes 8 bits to a 64-bit word, necessitating 8 chips per rank. All ranks within a channel share the same memory bus, allowing only one rank to be active at any given moment. Typically, a channel accommodates up to 4 ranks.

Each DRAM chip is composed of several banks, which can be addressed individually. However, since they share the same datapath on the DRAM chip, only one bank can be actively transferring data  at any given time. To maximize performance, multiple commands to different banks are often pipelined together, allowing for bank-level parallelism, for example by reading from one bank while precharging another.
The terminology associated with banks can be confusing. The logical concept of a bank refers to an independent division of memory that is striped across all the chips within a rank. Meanwhile, a physical bank refers to a a single chip's portion of this rank-wide bank. Some works use the term ``sub-bank" \cite{memory-systems-synth-lecture} to refer to the physical concept of a bank. 
For the purposes of this paper, the term ``bank-level" refers to a single bank within a specific chip. A typical rank can consist of 8-16 logical banks, i.e. each DRAM chip has 8-16 physical banks.

\begin{figure}
    \centering
    \includegraphics[width=0.48\textwidth, bb=0.410766 4.449797 512.514828 187.383299]{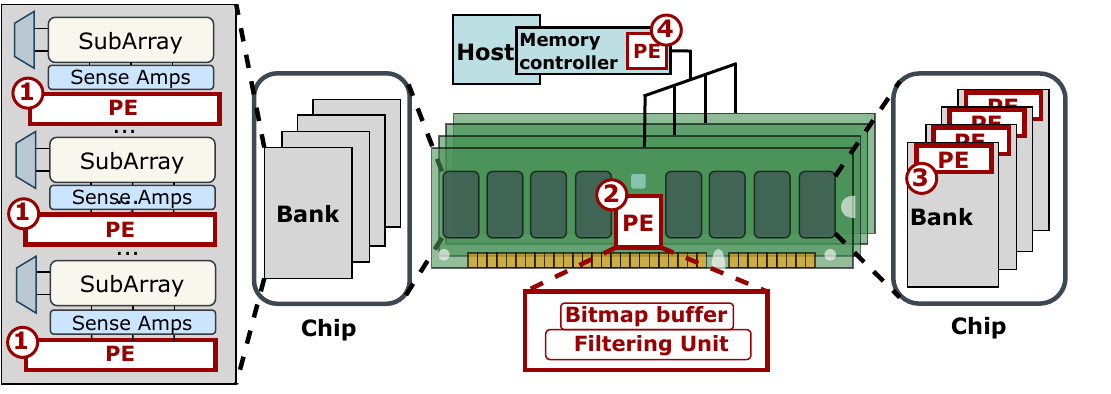}
    \caption{Membrane Design Space Exploration: {\large{\ding{172}}} Processing Element (PE) at the Subarray-level {\large{\ding{173}}} PE at the Rank-level {\large{\ding{174}}} PE at the Bank-level {\large{\ding{175}}} PE at the Channel-level}
    \label{fig:membrane_dse}
\end{figure}

The physical banks are divided into subarrays; several bits from the row address select the subarray, and the remaining bits select the row within the subarray. Each subarray has its own row and column decoders, as well as peripheral circuitry such as a subarray-wide row buffers. While multiple subarrays can be considered independent, only one can be addressed at a time, due to the common datapath they share, which consists of shared row-address and column-index buses and a shared global data line (GDL) for sending the selected word to the bank interface. In a typical access, the desired bank must first be precharged, which sets the bitlines to Vdd/2.  After the row has been read from the selected subarray into the sense amps, which amplifies the analog change in the bitline relative to a reference, the column decoder selects the desired word and places it onto the GDL, which brings the word to the bank's global row buffer. From here it is placed on the chip's I/O pins.  Figure \ref{fig:membrane_dse} shows this hierarchy and some of the locations where we insert near-data or PIM logic, which will be discussed further in \Cref{sec:PIM_architectures}. Note that in open-page mode, the DRAM holds the values in the row buffer, so subsequent reads to the same row do not need to precharge and can fetch data from the subarray's row buffer again; the latency between successive reads from the same row is $t_{CCD}$, on the order of 4--8 DRAM clock cycles, while accessing a different row requires waiting for precharge and waiting for the sense amp values to settle, i.e. $t_{RAS}$ + $t_{RCD}$ + $t_{CL}$, on the order of approximately 100 cycles.  

\section{Mapping Data Analytics to PIM}


\begin{table}
\small
\begin{center}
\begin{tabularx}{0.5\textwidth}{|X|c|c|c|} 
    \hline
    PIM-Amenability Test & Filter & \makecell[t]{Aggregate \\(sorting)} & JOINs \\ [0.0ex] 
    \hline\hline
    Memory-bound? & \textcolor{teal}{\ding{52}} & \textcolor{teal}{\ding{52}} & \textcolor{teal}{\ding{52}} \\ 
    \hline
    Low cache-reuse? & \textcolor{teal}{\ding{52}} & \textcolor{red}{\ding{56}} & \textcolor{red}{\ding{56}} \\
    \hline
    Localized operand interaction? & \textcolor{teal}{\ding{52}} & \textcolor{red}{\ding{56}} & \textcolor{red}{\ding{56}} \\
    \hline
    Aligned Data Parallelism? & \textcolor{teal}{\ding{52}} & \textcolor{red}{\ding{56}} & \textcolor{red}{\ding{56}} \\
    \hline
    Run on $\rightarrow$ & \textcolor{teal}{PIM} & \textcolor{red}{CPU} & \textcolor{red}{CPU} \\ [0.0ex] 
    \hline   
\end{tabularx}
\end{center}
\caption{ Major kernels used in Analytics Database Workloads and their PIM Amenability characteristics. }
\label{table:pim_tests}
\end{table}

\subsection{PIM architecture requirements for data analytics}
The landscape of near-data and in-memory processing is extensive, and our principal objective is to optimize specifically for {\em in-memory} analytics workloads. 
Given that the data are already in memory, we seek a solution that can operate on the data in place, allowing processing in memory and regular load/store access to the same data, without the need to move data between PIM-friendly and CPU-friendly layouts. 
We also seek a solution with sufficiently low overhead to justify inclusion within a commodity (albeit premium) DRAM product.  

While many PIM designs have been proposed in the past, we look for inspiration to three major commercially-announced PIM systems that respect the constraints mentioned earlier: Samsung's Aquabolt (implemented in HBM) \cite{aquabolt-isca-2021}, SK hynix's AiM (based on GDDR) \cite{sk-hynix-aim}, and UPMEM (DDR) \cite{upmem-benchmarking}. We collectively refer to the Aquabolt and AiM architectures, which share significant similarities, as HBM-PIM.

All three extant PIM products place logic at the bank interface, but with significant hardware overhead. Aquabolt and AiM target acceleration of neural network kernels such as  GEMV and support SIMD arithmetic units at the bank interface, with significant impact on capacity per unit area, 50\% and 20-25\% respectively. However, servers designed for in-memory databases seem unlikely to adopt HBM, and currently use traditional SDRAM DIMMs, because this technology is much lower cost and scales much more easily to the sizes needed. UPMEM on the other hand implements independent tasklet-based processing at the DDR's bank level, using a 64KB scratchpad per bank and data processing units (DPUs) that can operate independently, in a task-parallel manner. We have not been able to find information about the area overhead of this approach.

We also considered subarray-level PIM (placing units at some or all of the subarrays in each of the banks), rank-level near-data processing (placing units on the DIMM module, not in the DRAM chips) and channel-level filtering (placing units at the channel interface just before the cache hierarchy) and will show that the bank-level approach provides the best balance of performance vs.\ overhead.  

\subsection{PIM Amenability Tests}
Prior work \cite{inclusive-pim} studies the HBM-PIM \cite{aquabolt-isca-2021} style architecture and proposes four PIM-Amenability Tests to assess whether a kernel is suitable for PIM acceleration. The work proposes that a kernel should pass all four tests and not just a subset of them to be well-suited for PIM.  Table~\ref{table:pim_tests} shows how the three major stages of database analytics, filter, joins, and aggregation, map to these criteria.  Only filtering meets all four criteria.

The four major criteria that the test suggests are as follows:

\begin{enumerate}
\item Is the workload memory-bound? Bandwidth inside the DRAM is much higher than the bandwidth of the DRAM interface.  If the workload is memory bound and can effectively use this higher internal bandwidth, then it is suitable for PIM acceleration. Otherwise, PIM may save energy but is less likely to boost performance.  

\item Does the workload have low cache reuse? If not, better performance is typically achieved via CPU computation, which operates at a much higher clock speed and benefits from cache reuse. 

\item Are computations localized within a single bank? Transfers between banks or ranks are costly. 

\item Does the workload exhibit memory-aligned data parallelism? PIM architectures that leverage bank and/or subarray-level parallelism compute simultaneously on the same row and column addresses across banks/subarrays. Thus, data must be aligned to be executed in lockstep across multiple banks.  Furthermore, this type of data parallelism maximizes row buffer locality.
\end{enumerate}

We would suggest another consideration related to item 4 above: Does the algorithm exhibit sufficient parallelism and operate on large enough data objects to leverage sufficient internal parallelism of the DRAM to show speedup over near-data processing outside the DRAM?

The filter kernel is memory-bound, because it does not exhibit temporal locality: it streams through the entire table, and elements that do not match the filter predicate are not touched again. Column-oriented schema do exhibit spatial locality, but because the computation density per word is low (just a simple comparison), memory access remains the bottleneck. Typical in-memory databases are many GB in size, and filtering is also embarrassingly parallelizable, allowing full use of the DRAM's internal parallelism, and filtering exhibits aligned data parallelism. 



Joins, although memory-bound \cite{join-roofline-plot}, benefit from having caches while performing hash-join. The join algorithms are tweaked in many instances \cite{blanas2011design,balkesen2013main,schuh2016experimental} to make the join algorithm cache-aware, and in many cases the dimension tables used to create the hash table for the hash join are small enough to fit easily in the CPU cache hierarchy. We also observe that doing a hash join in DRAM would likely require a copy of the hash table in each bank to avoid cross-bank interactions, although this could also be stored in the bank itself, and hashing typically entails random accesses to the hash table, inhibiting aligned data parallelism. Hence, join kernels do not meet 3 out of the 4 PIM-amenability criteria. Furthermore, in comparison to join, filtering requires only a simple comparison per predicate, instead of hashing plus table lookup, and denormalization is able to convert joins to simple, PIM-friendly, streaming comparison operations, without the complexity of hashing.




Aggregation (grouping, sorting, etc.) is only performed on the selected records. It exhibits greater temporal cache locality, and involves more complex computation that would be difficult to localize within a bank and would require more costly processing units in the PIM.

Our results, shown in the upcoming sections, show that bank-level filtering units are so effective that they reduce time spent on filtering to a negligible proportion of execution time, and minimize the amount of data that subsequently needs to be fetched by the CPU.  This approach is so effective that a more aggressive PIM approach, such as subarray-level PIM, rarely provides meaningful additional end-to-end performance benefit---with bank-level PIM, the filtering step is already reduced to such a small portion of the overall execution time that further hardware cost to achieve greater speedup is not worth the additional hardware cost.  But in comparison to channel- or rank-level processing, the bank-level approach provides significant speedup, with  tiny hardware cost.


Once the filtering kernel has produced its output bitmap, the rest of the query is processed on the CPU. The necessary fields from only the selected records, as indicate by the bitmap, are fetched to the CPU.



\section{PIM Architectures}\label{sec:PIM_architectures}



\begin{figure}
    \centering    \includegraphics[width=0.5\textwidth, bb=0.850430 3.906000 601.914989 167.819831]{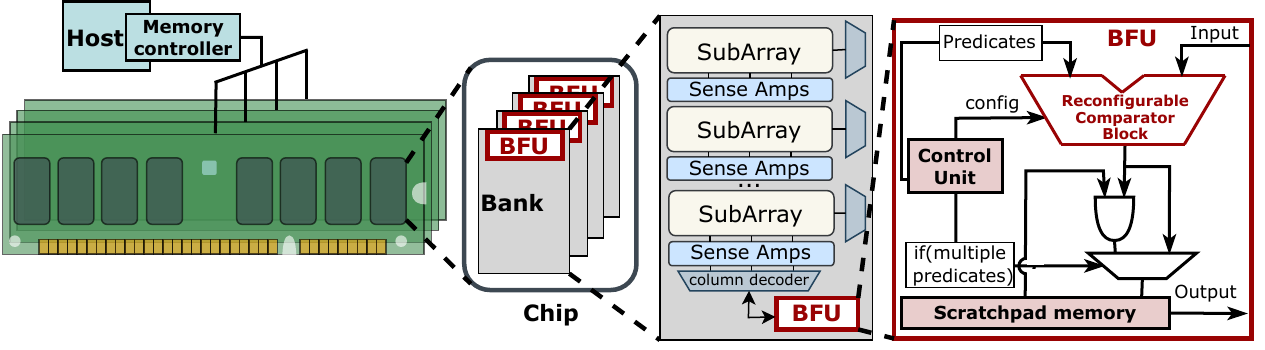}
    \caption{System with Membrane Bank-level Filtering Unit}
\label{fig:DIMMs_with_Bank_architecture}
\end{figure}

\subsection{Bank-level Filtering Unit (BFU)} \label{sec:bfu}
Our proposed Bank-level Filtering Unit (BFU) only needs to support comparisons for the filtering step. The BFU is placed at each bank interface, which fetches 64 bits in a burst from the subarray row buffer. The BFU processes data every tCCD\_L in All-Bank mode, like Aquabolt \cite{aquabolt-isca-2021}.  The breakdown of BFU is presented in \Cref{fig:DIMMs_with_Bank_architecture}, and its components are described below.

The BFU's \textbf{Reconfigurable Comparator Block (RCB)} can support both equality check (database\_value==a) and range check (a\textless database\_value\textless b) on integer and floating point values. Each comparison within BFU produces a result bit that is placed into a bitmap stored in the BFU's bitmap buffer, which is 64 bits. When performing a sequence of multiple filtering operations, the RCB reads the value of the bitmap for the current position and ANDs this with the result from the new comparison operation. This way, the bitmap accumulates the final Boolean result for an entire query. Once the 64-bit output buffer is filled, it is written back. The RCB does not support processing string or regex operations, as these are infrequent and more costly for PIM implementation. In any case, strings would be dictionary-encoded. 

Modern databases store columnar data in a bit-packed format to decrease memory usage, so that fields with a small range, such as zip codes, do not waste space.  Our comparison unit supports any bit length from 2 to 64 bits, with smaller bit widths processed in SIMD fashion. Configuration bits indicate the data type to be processed. Thus, before processing a database column, the RCBs are configured to the specified bit length, programmed with the predicate values to compare against, and then the filtering operation begins to produce the desired resultant bitmap.  The \textbf{Control Unit} orchestrates fetching data from the DRAM bank and performing the comparison at the desired bit lengths. 

\subsection{Subarray-level Filtering}

To explore subarray-level parallelism, we adapt the Fulcrum architecture~\cite{fulcrum}, which places a PE at the edge of the subarray, and uses Walkers (row-wide buffers) and column-select logic to move input operands out to the PE, and move  output values to the appropriate location in the output Walker.  Due to the open-bitline architecture, we can have at most one PE for every two subarrays. The full Fulcrum architecture supports arithmetic, etc., while filtering only needs comparison, and only needs two  Walkers to read the input column values and capture the resulting bitmap values. Our subarray-level PE is similar to the BFU, with a Walker used to hold the bitmap. The second Walker means the area overhead per filtering unit is significantly higher than the bank-level approach.   For our chosen DRAM configuration (\Cref{tab:workstation}), CACTI \cite{cacti} indicates 16 subarrays per bank with each subarray containing 4096 rows. 
For 16 subarrays, the maximum subarray-level parallelism (SALP) is at most 8, because each PE is shared by two subarrays.  For smaller degrees of SALP, data may need to be moved from a subarray that does not have access to a PE to one that does using the LISA technique \cite{lisa}.

As shown in \Cref{fig:scan_microbenchmark}, subarray-level PIM provides minimal extra performance compared to the bank-level approach, at the cost of higher area, so we do not consider further design optimizations.

\subsection{Rank-level and Channel-level Filtering}
Placing computation outside the DRAM on the DIMM module or in the memory controller avoids changes to the DRAM but also gives up the higher internal parallelism of the DRAM. We explore the rank-level filtering by placing a  BFU in a buffer chip on the DIMM circuit board. This rank-level filtering unit can process an entire 64-bit DRAM read burst in one cycle, and is an upper bound for filtering throughput outside the DRAM chips if we maintain a standard-width DRAM interface. For the channel-level filtering, we place a similar unit in the memory controller. This channel-level filtering unit provides a rough approximation of what the Intel Analytics Accelerator (IAA)~\cite{intel-iaa} can achieve, by offloading filtering from the cores and avoiding cache pollution. We model the channel-level filtering in this way, like the rank-level filter unit, for better comparison with the rank-level approach, and  because we were not able to find specific implementation details of the IAA.  The only significant difference between our channel- and rank-level filtering is  that the rank-level approach has one unit per rank, thus achieving rank-level parallelism.  

As our results show, the speedup at the bank level, compared to rank- and channel-level, is substantial (proportional to the number of banks), so we do not consider further optimizations for rank- and channel-level processing.

\subsection{System Integration}
\label{sec:architecture-system-integration}

Following the Aquabolt approach, which maintains compatibility with existing DRAM interfaces, Membrane supports Single-Bank (SB) mode (normal read/write) and All-Bank (AB) mode for PIM. In AB Mode, a DRAM READ command to a specific address reads data at the address into the local BFU and performs a PIM computation (i.e., comparison). In AB mode, the bank and bank-groups bits in a given memory address are ignored, and data at the same columns position across all the banks is read into the local BFUs.  As with Aquabolt, Membrane uses MRS (Mode Register Switch) and PIMCONF (PIM Configuration) registers to control the functionality of PIM processing elements.  MRS  is used to transition between the normal mode (SB mode) and PIM-capable mode (AB Mode).  PIMCONF registers are used to program the PIM processing elements with instructions. In our case, we use the PIMCONF registers to program the BFU with the values to be compared against during the predicate operations and set the processing bitwidth.

\begin{figure}
    \centering
    \includegraphics[width=\linewidth, bb=0.720211 1.669711 325.529779 38.399764]{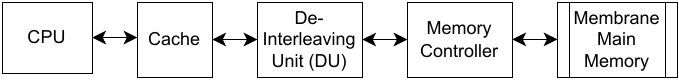}
    \caption{PIM System Integration with Cache De-interleaving Unit (DU)}
    \label{fig:system_integration}
\end{figure}

\textbf{Cache De-interleaving Unit (DU).} A cache line is read or written in 64-bit chunks, and a single 64-bit chunk of data is usually striped across multiple chips within a single DDR rank. Traditionally, DDR comes in x4, x8, or x16 configurations, in which each xN chip in the rank contributes 4, 8, or 16 bits to a 64-bit DRAM access. For example, in a x8 configuration, each memory chip holds 8 bits of a 64-bit word.  This striping across multiple chips is problematic for PIM when processing operands greater than the 4-, 8-, or 16-bit width, because different bytes of an operand are spread across multiple chips, preventing even simple comparisons. We change the interleaving slightly to support PIM, so that each 64-bit word is kept together in a single DRAM chip. However, reads/writes from the CPU memory controller still fetch 4, 8, or 16 bits from each DRAM chip, so the traditional memory controller read will bring in bytes that are not adjacent in the cache line.  This is easily addressed by adding a cache-line-wide buffer to each memory controller.  The memory controller is configured with the interleaving, so now, each DRAM read routes the incoming bits to the appropriate location in the buffer, and over the course of the eight reads needed to fill a cache line, all the bytes are filled in. For example, with x8, each byte in the 64-bit read from DRAM will go to a separate 64-bit word in this buffer, as shown in Figure~\ref{fig:system_integration}. Writing a cache line to memory reverses the process. The overhead for this buffer and routing the incoming data in this fashion is negligible.  


\textbf{PIM Pages.} When running in AB Mode, each DRAM command  triggers a READ and PIM operation across a single column in the same row across all banks. 
A {\em PIM page} is the enforced minimum allocation unit for PIM, and is a multiple of the native operating system superpage size. The PIM page size will depend on the system's configuration, so the PIM page fills at least one entire system-wide DRAM row, i.e. spanning this row ``position" across all banks, ranks, and channels.  For a large memory system, this will require multiple contiguous superpages, e.g., for an 8-channel, four ranks/channel system with DDR4\_8Gb\_x8\_2933 DRAM chip configuration taken from \cite{dramsim3}, a PIM page is 4 MByte, requiring two 2 MByte superpages on an x86-64 system.   For a smaller system, a single superpage will suffice, and it might occupy several logical rows. This means that when building a system to use Membrane memory, the memory system should configure memory channels in powers of 2, and if the data cannot fill a PIM page, it should be padded. This may also requires another minor change to the typical address interleaving, so that PIM pages use full rows in the DRAM.

Using superpages allows us to allocate PIM data with permissions enabling PIM.  The operating system must support a new version of malloc, filter\_malloc(), that gives the owner permission to issue PIM commands to this region of memory. A filter\_malloc is needed for each input PIM page as well as each output PIM page, for storing the output bitmap. PIM permission must be noted in the page table and requires one extra bit in the TLB. The system must also support virtual-to-physical translation and permission checking at the granularity of PIM pages; more on this in the next subsection. No other changes are required to the operating system or MMU. Leveraging the superpage feature benefits from the reduced TLB lookups provided by superpages. Note also that PIM pages do not need to require any particular placement in memory or relative to each other.

\begin{figure}
    \centering
    \includegraphics[width=0.6\linewidth, bb=6.098555 1.944000 198.683994 139.211996]{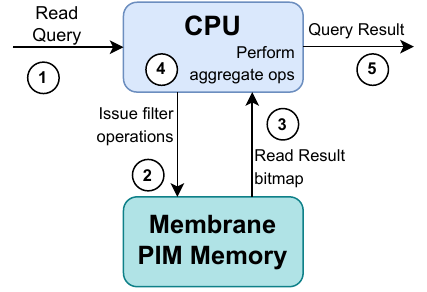}
    \caption{Query Execution Flow with Membrane }
    \label{fig:execution_flow}
\end{figure}

\textbf{Query Processing and Additional System/Hardware Support}
Membrane requires several new CPU instructions to control PIM operation, as noted below. The description below is specific to bank-level PIM, but can easily be adapted to subarray-level, etc.
The overall execution flow is shown in \Cref{fig:execution_flow}.
The software performs PIM filtering on a column by operating on one PIM page at a time.  We envision this is implemented in a PIM user-level filter library that any database engine can incorporate. Only one thread in the database  should perform PIM. 

The library first issues a  pim\_begin() system call. This allows the OS or hypervisor to manage access to PIM mode, and return an error if the calling process is not allowed to use PIM. It can also support fair sharing for PIM access, etc. 
Although Aquabolt allows user-level access to PIM mode, we suggest that access be mediated by the OS. The pim\_begin() system call writes into the MRS register to drain the memory controller queues and switch all channels into All Bank mode. This also blocks regular load-store access from other threads/cores. When the system call returns, the user-level  library  programs the BFUs with the predicate values and operation type using a PIM\_CONF instruction that writes to the PIMCONF address, which is broadcast to all banks' BFUs. 

Now filtering can begin using a sequence of PIM\_FILTER instructions, each specifying one PIM input page, one PIM output page, and an offset within the output page (because input values may be up to 64 bits but the corresponding output value is just a single bit, so processing an entire input PIM page only fills up a small portion of the output page).  These addresses are translated through the TLB to obtain the physical address for the PIM page and verify PIM permission, and these physical addresses are sent to all memory controllers. For each PIM\_FILTER instruction, the memory controllers process their portion of the PIM page by sending the appropriate sequence of PIM DRAM commands to activate the target DRAM row and sequence through this row's DRAM columns.  The latency for each of these DRAM commands is deterministic, so the memory controller knows how long to wait before initiating a subsequent command. A PIM\_FILTER instruction is issued for each PIM page needed to complete the filtering required by the column. Once filtering the database column is complete, a new column can be processed. When PIM computation is complete, the library issues a pim\_end() system call, which reverts the memory to standard operation. Note that we do not attempt to offload the PIM computation from the core, because AB mode blocks the entire memory system anyway, and the calling thread is waiting on the PIM results. In summary, PIM is not completely transparent to the software. The database application needs to allocate the data in memory using pim\_malloc(), and use the PIM-enabled filter library.

Filtering is idempotent, so if a non-maskable interrupt occurs that requires DRAM access, for simplicity, the filter operation can be aborted and restarted later. Any prior partial bitmap results will be recomputed. This avoids the need to preserve partial PIM state.

For other interrupts, the OS can wait until a PIM page has been completed, then transition out of PIM mode if needed.  If the process performing PIM computation needs to be suspended, the OS only needs to remember to re-initiate PIM mode when this process resumes.  The process's progress in the filtering task is remembered in its user state.

The overhead of setting up  PIM mode is a system call and the DRAM writes to set up PIM mode, plus the time to finish any pending memory operations, and the overhead for exiting is another system call. The minimum Linux system call latency is on the order of 1-2 ms. This latency is negligible compared to a sufficiently long sequence of PIM\_FILTER commands, each of which performs $rowsize/64 bits$ column accesses per DRAM row, with a latency of $0.38 \mu s$. We confirm this by assuming the standard DDR4 setup containing 128 columns per bank and modelling the all-bank mode in DRAMSim3. For a minimum OS timeslice (sysctl\_sched\_min\_granularity) of $0.75 ms$, this allows at least 1968 PIM\_FILTER operations. For SSB scale factor 100, one column of, e.g. 32-bit values occupies 573 PIM pages. 

We assume PIM is used in a system that is primarily supporting a data analytics workload, so that blocking the memory system for the duration of an OS timeslice is acceptable. If the system operator wants to reduce the amount of time the memory system is blocked, a different minimum time slice can be used specifically for PIM mode. 

No changes are needed to support multi-tenancy, since each VM will have its own memory allocations. The hypervisor will need to support PIM mode, and the OS overhead for initiating PIM mode would increase. 

DRAM refresh is managed by the memory controller.  Each row is refreshed every 64 millisec, so the memory controller can issue refresh operations when needed between PIM\_FILTER operations.

Once all filtering is completed for the query, the database software now reads the bitmaps from the main memory, and accesses any needed fields for records that have been selected. The aggregate operations (sorting, group by, average or as such) to produce the final result.

\section{Experimental Methodology} \label{sec:methodology}


\textbf{DRAM simulation and host system.} In Membrane, filters are executed in PIM, and the remaining work of the query is executed on the host CPU, taking the bitmap as input. To model the PIM portion, we use DRAMSim3 \cite{dramsim3}, a cycle-accurate DRAM simulator. When a PIM page is activated, the entire page is brought into the row buffers across all the banks and all the ranks, and then each bank processes its portion in 64-bit chunks. The precharge, row activation, and reads are modeled in DRAMsim3 to obtain the time required to filter an entire PIM page in AB mode. To evaluate the host CPU portion, we use DuckDB, a state-of-the-art OLAP database system \cite{duckDB}. DuckDB is extensively optimized, outperforming more established systems by orders of magnitude in many cases, and is widely used as a baseline in database research \cite{duckdb-baseline-1,duckdb-baseline-2,duckdb-baseline-3,duckdb-baseline-4}. The total end-to-end time spent processing a query is the sum of time spent on PIM filters (simulated) and the rest of the query in DuckDB (real-world execution). Note that, to avoid iterating through sparse bitmaps, we modify DuckDB to leverage CPU instructions that count the number of trailing zeroes in a word.



\textbf{Workload.} To evaluate the many dimensions of the DRAM-PIM filtering design space, we use two established OLAP benchmarks: TPC-H \cite{tpch} and the Star Schema Benchmark (SSB) \cite{ssbbenchmark}. TPC-H is a widely-used OLAP benchmark designed to comprehensively assess the performance of OLAP systems. The TPC-H database consists of a large, central \texttt{lineitem} table emulating the items ordered from a business. Dimension tables store information about parts, suppliers, customers, and locations. The benchmark consists of 22 queries. To focus our evaluation, we use a subset of 8 queries that have been used in prior work focused on evaluating filter performance \cite{sun2014fine}. We report the geometric mean of this subset to summarize our results. While SSB is based on TPC-H, it includes notable distinctions that aim to improve its accuracy and coverage as a benchmark. SSB combines the \texttt{lineitem} and \texttt{orders} tables, a standard technique used to avoid unnecessary joins \cite{kimball2002}. It also drops tables and columns that are unlikely to be present in an OLAP database, such as string comments and shipping instructions. The query suite consists of 4 ``flights", each of which models a common OLAP pattern. Within each query flight, there are several queries with varying selectivity (the number of rows that contribute to the result of each query). The database itself consists of one large fact table and four smaller dimension tables.

\begin{table}[t]
\scriptsize
\centering
\caption{\centering{Configuration Details.}}
\begin{tabular}{|c|c|}
\hline
Property                  & Value \\
\hline
\hline
Baseline System            & Intel Xeon Silver 4314 @ 2.40 GHz \\
Total Cores / Main Memory              & 16 Cores (32 threads) / 128 GB (8-ch DDR4-3200) \\
\hline
PIM Config.            &   DDR4\_8Gb\_x8\_3200     \\  
                            &   8-channel, 4-rank/ch, 4-bank-group                  \\
                            &   4 banks per bank-group, 16 subarrays/bank \\ 
                            
                            \hline
\end{tabular}
\label{tab:workstation}
\end{table}

\textbf{Membrane Circuit Evaluation.} We implement Membrane's Bank-level Filtering Unit in RTL and use  Synopsis DC Compiler in 14 nm to evaluate its delay, power, and area. We use scaling factors from Stillmaker et al.~\cite{cmosscaling} to scale down the results to 22nm. Each BFU occupies $0.001{mm}^2$ area, which is negligible, and has a path delay of $0.45ns$, which easily fits within the column-to-column access time. The power for one BFU is $118.7uW$, which when aggregated across all banks within a rank is 2.3\% more than the regular DRAM operation.   

However, AB mode operates all banks at once, which increases peak power. 
Another PIM architecuture \cite{newton} that leverages the AB mode observes that the peak power increases by 4x when operating in this mode. Our evaluations consider these increased power requirements, and we observe a 3.6/3.1x relative energy efficiency over a baseline system without Membrane for SSB/TPCH benchmarks, respectively.


\textbf{Energy Consumption Analysis.} 
We estimate the power consumption of CPU while performing filter and non filter kernels based on CPU usage using the methodology in \cite{power_cpu_usage}. The overall energy consumption is obtained by integrating CPU power, DRAM power (obtained from DRAMsim3), and BFU power (obtained from the RTL analysis above) over the time spent on the filter and non filter kernels of end-to-end query execution. In AB mode, the extra power is included for the duration of PIM execution.  The relative energy efficiency highly correlates with the end-to-end execution time of the queries. We observe higher energy efficiency ($\sim$20x) with more selective queries such as Q3.3, Q3.4, Q19.


\label{sec:methodology}


\begin{table}[t]
\scriptsize
\centering
\caption{\centering{Levels of Schema Denormalization.}}
\begin{tabular}{|c|l|}
\hline
Denorm. Level & Description                              \\ 
\hline\hline
D1 & No Denormalization (Plain Schema) \\
\hline
D2 & Denormalizing columns used in where clause only \\
\hline
D3 & D2 + Columns used for aggregate operations \\
\hline
D4 & Full Schema Denormalization \\ \hline
\end{tabular}
\label{tab:denorm_explanation}
\end{table}

\begin{figure}
    \centering
    \includegraphics[width=2.5in]{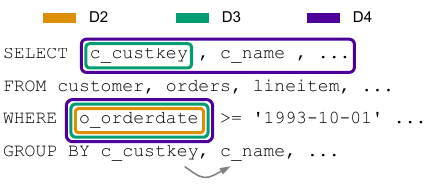}
    \caption{Illustrating which columns are denormalized in each level with TPC-H Q10. In D3, \texttt{c\_name} is not denormalized because it is functionally determined by \texttt{c\_custkey}.}
    \label{fig:example-query}
\end{figure}





\textbf{Denormalization.} To improve PIM amenability and fully exploit Membrane's capabilities, we explore the use of denormalization. As introduced in \Cref{sec:bkg:DB}, denormalization involves joining tables as the database is loaded. Commonly used to reduce query complexity and improve performance, denormalization replaces joins with filters. However, denormalization requires extra space to store the denormalized data.

The choice of which columns to denormalize presents a tradeoff between PIM amenability and space overhead, as shown in Table~\ref{tab:denorm_explanation}. At one extreme (D1), we can avoid denormalization, which incurs no space overhead but limits the speedup. At the other extreme (D4), we can denormalize all columns, which maximizes PIM amenability at the expense of considerable space overhead. As reported in prior work, full denormalization can result in a space blowup of over 10x.

\begin{figure*}
    \centering
    \begin{subfigure}{3.3in}
        \includegraphics[width=\linewidth]{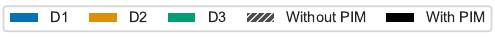}
    \end{subfigure}
    
    \begin{subfigure}{4.12in}
        \includegraphics[width=\linewidth]{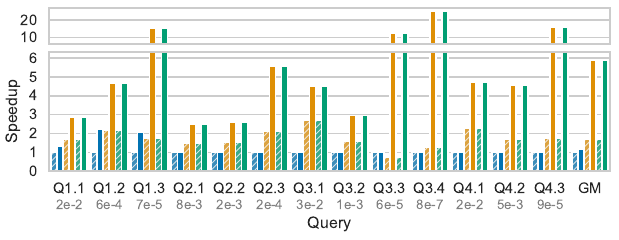}
        \caption{SSB}
        \label{fig:speedup-vs-query-and-schema-variant-ssb}
    \end{subfigure}
    \hfill
    \begin{subfigure}{2.78in}
        \includegraphics[width=\linewidth]{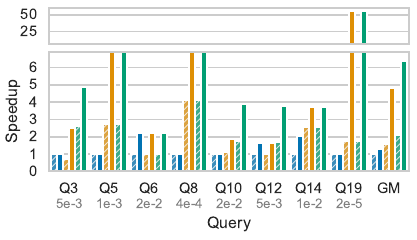}
        \caption{TPC-H}
        \label{fig:speedup-vs-query-and-schema-variant-tpch}
    \end{subfigure}
    \caption{Query speedup for varying denormalization level (relative to D1 without PIM). Query selectivity is shown at the bottom.}
    \label{fig:speedup-vs-query-and-schema-variant}
\end{figure*}

We propose two denormalization levels, D2 and D3, which fall between the two extremes, offering a better balance between PIM amenability and space overhead. Inspired by WideTable \cite{widetable}, we use static analysis of the workload to choose a subset of columns to denormalize. In addition, we use dictionary encoding and bitpacking compression in all our experiments to reduce space overhead, which is particularly beneficial for denormalization.  These techniques do not affect PIM amenability. 

In D2, we denormalize a column if it appears in the \texttt{WHERE} clause of any query in the workload. Recall from \Cref{sec:bkg:DB} that in D1, rows of interest are selected through a combination of filters and joins. D2 replaces these joins with filters.

D3 is motivated by the observation that a significant portion of query time is spent on joins even after denormalizing columns that appear in the \texttt{WHERE} clause, as shown by the D2 breakdown in \Cref{fig:operator-time-percentage-vs-schema-variant-tpch}. For certain queries, joins are used to retrieve columns that do not appear in the \texttt{WHERE} clause but are still needed to answer the query. In D3, we reduce the impact of these joins by denormalizing a column if it appears in the \texttt{WHERE} clause or the \texttt{SELECT} clause of any query in the workload. To reduce space, D3 involves a notable exception: we do not denormalize dimension table columns that only appear in the \texttt{SELECT} clause and are functionally determined by another column in the \texttt{GROUP BY} clause. The exception is based on the observation that group-by aggregation and limit operations often reduce the number of rows down to the order of tens to hundreds. After these operations have been completed, an inexpensive join can be used to retrieve the functionally dependent columns and produce the result. As illustrated in \Cref{fig:example-query}, D3 avoids denormalizing \texttt{c\_name} and other large columns from the \texttt{customer} table. 


\section{Results}
\subsection{Filter Performance Across the DRAM Hierarchy} \label{sec:scan_microbenchmark}

We previously explained why the filter kernel is the most suitable candidate for PIM acceleration and suggested that the bank level is the best choice for adding PIM computation for filtering. Now we substantiate this claim by briefly evaluating the benefits of placing PIM filtering at different levels of the DRAM hierarchy. 

To assess the benefit of filtering at different levels of the DRAM hierarchy, we constructed a microbenchmark that  performs a simple predicate (a\textless input\_value\textless b) evaluation on one column of the Star Schema Benchmark's (SSB) fact table. Each fact table column at scale factor-100 contains 600,038,146 elements; for this microbenchmark, the values are 16-bits each (for a total of 1.12 GB). 

\begin{figure}
    \centering
    \begin{subfigure}{2.2in}
         \includegraphics{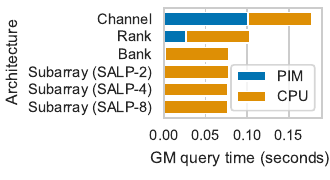}
    \end{subfigure}
    \hfill
    \begin{subfigure}{1.2in}
        \includegraphics{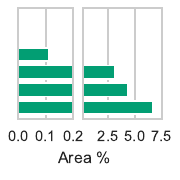}
    \end{subfigure}
    \caption{Geometric mean SSB query time and area overhead (relative to cell area) for varying PIM architectures.}
    \label{fig:scan_microbenchmark}
\end{figure}

\Cref{tab:scan_microbenchmark} shows the latency for our microbenchmark with different forms of near/in-memory processing. While 8-way subarray-level parallelism is able to achieve 14x speedup over the bank-level approach on our microbenchmark, when considering geometric-mean end-to-end performance of the SSB and TPC-H suites, as shown in \Cref{fig:scan_microbenchmark} along with area overhead, the speedup advantage with SALP over baseline CPU drops to 1.1x speedup (SALP-8 vs Bank-Level), which does not appear to justify the much higher area cost.

Comparing between rank-level and bank-level, we observe that rank-level PIM does not have any area overhead inside the DRAM chips, but it is 29.4x slower than the bank level approach in the microbenchmark. However, when considering the geometric-mean end-to-end performance of the SSB/TPC-H suites, the bank-level solution offers 1.89x/1.59x speedup over rank-level with 4 ranks/channel. 

Filtering could also be performed in the memory controller or some other unit in the CPU, as in the Intel Analytics Accelerator~\cite{intel-iaa}, which offloads this data-intensive task from the cores and avoids cache pollution, but gives up the rank-level parallelism of the rank-level solution. Bank-level offers 3.7x/3.15x speedup (SSB/TPCH) over this channel-level solution. 

Based on these findings, we conclude that the bank is the best level of the DRAM hierarchy in which to implement filtering, with only 0.1\% area overhead relative to the baseline DRAM chip area. 

\begin{table}[t]
\centering 
\caption{Single-column filter latency} 
\label{tab:sampletable} 
\begin{tabular}{|c|c|c|c|c|c|c|} 
\hline 
PIM Arch. & Chnl & Rank & Bank & SALP-2 & SALP-4 & SALP-8\\ \hline
Time (ms) & 32.4 & 8.46 & 0.28 & 0.08 & 0.04 & 0.02 \\ \hline
\end{tabular}
\label{tab:scan_microbenchmark}
\end{table}

\subsection{Partial denormalization enables more extensive acceleration}

We evaluated the overall performance of Membrane bank-level PIM's performance with SSB and TPC-H benchmarks against the baseline system configuration (\Cref{tab:workstation}). Speedups directly correlate with query selectivity. We observe that with the accelerated PIM filters, we obtain end-to-end geo-mean query speedup of 5.92x/6.38x in SSB/TPC-H while using the D3 schema, but only 1.2x and 1.3x for SSB and TPC-H if denormalization is not used (D1). 

\begin{figure}
    \begin{subfigure}{1.1in}
        \includegraphics{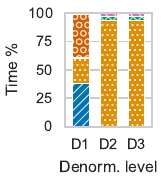}
        \caption{SSB}
        \label{fig:operator-time-percentage-vs-schema-variant-ssb}
    \end{subfigure}
    \hfill
    \begin{subfigure}{1.1in}
        \includegraphics{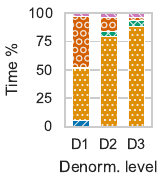}
        \caption{TPC-H}
        \label{fig:operator-time-percentage-vs-schema-variant-tpch}
    \end{subfigure}
    \hfill
    \begin{subfigure}{1.1in}
        \includegraphics{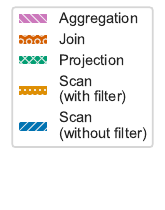}
    \end{subfigure}
    \caption{Average operator time percentage for varying denormalization level (without PIM).}
    \label{fig:operator-time-percentage-vs-schema-variant}
\end{figure}

To better understand the benefits of denormalization, in \Cref{fig:operator-time-percentage-vs-schema-variant}, we show the average percentage of time spent in each operator for SSB and TPC-H without PIM acceleration. For denormalization level D1 (the standard schema), scans (both with and without filtering) account for 60\% and 51\% of query time in SSB and TPC-H. As shown in Figure~\ref{fig:speedup-vs-query-and-schema-variant}, for denormalization level D1, Membrane
achieves over 2x speedup for SSB Q1.2 and Q1.3 and TPC-H Q6 and Q14, which are dominated by filtering. Unfortunately,
because scans with filters account for only 22\% of overall SSB query time and 46\% of overall TPC-H query time, Amdahl’s law limits the overall potential speedup to approximately 1.3x and 1.9x. However, \Cref{fig:operator-time-percentage-vs-schema-variant} shows that a substantial portion of time in D1 is also spent on joins, which dominate after D1 filtering is accelerated by PIM.

\begin{figure}
\centering
    \begin{subfigure}{1.1in}
        \includegraphics[width=\linewidth]{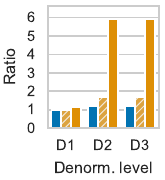}
        \caption{SSB}
        \label{fig:memory-overhead-and-speedup-vs-schema-variant-ssb}
    \end{subfigure}
    \hfill
    \begin{subfigure}{1.1in}
        \includegraphics[width=\linewidth]{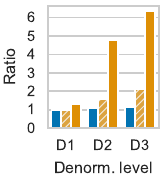}
        \caption{TPC-H}
        \label{fig:memory-overhead-and-speedup-vs-schema-variant-tpch}
    \end{subfigure}
    \hfill
    \begin{subfigure}{1in}
         \includegraphics[width=\linewidth]{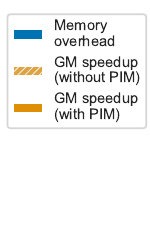}
    \end{subfigure}
    \caption{Memory overhead and geometric mean query speedup for varying denormalization level (relative to D1 without PIM).}
    \label{fig:memory-overhead-and-speedup-vs-schema-variant}
\end{figure}

At the expense of 17\% and 9\% extra space, as shown in Figure~\ref{fig:memory-overhead-and-speedup-vs-schema-variant}, D2 yields geometric mean query speedups of 5.9x and 4.8x
for SSB and TPC-H. Denormalization without
Membrane acceleration also improves performance, but to a
much lesser extent.  D3 further increases the portion of query time
that Membrane can accelerate. At the
expense of 3\% extra space, D3 achieves a geometric mean
query speedup of 6.4x for TPC-H. For SSB, D2 and D3
happen to be equivalent.  \Cref{fig:operator-time-percentage-vs-schema-variant} shows that with D3, joins have been nearly eliminated and converted to filters, and most of the execution time has been converted to filters.

\begin{table}[t]
    \centering
    \begin{tabular}{rrrrr}
    \hline
          & SF 10 & SF 20 & SF 50 & SF 100 \\
    \hline
    SSB   & 6.8   & 8.6   & 14.4  & 23.5   \\
    TPC-H & 7.8   & 10.8  & 21.7  & 39.3   \\
    \hline
    \end{tabular}
    \caption{Database size in GB for varying scale factor.}
    \label{tab:database-size-vs-scale-factor}
\end{table}

\begin{figure}
    \centering
    \begin{subfigure}{3.3in}
        \includegraphics{Membrane/Figures/speedup-vs-query-and-schema-variant-legend.pdf}
    \end{subfigure}

    \begin{subfigure}{1.6in}
        \includegraphics{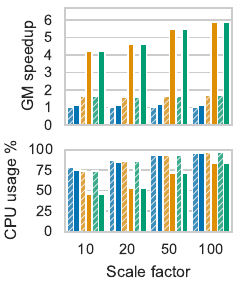}
        \caption{SSB}
    \end{subfigure}
    \hfill
    \begin{subfigure}{1.6in}
        \includegraphics{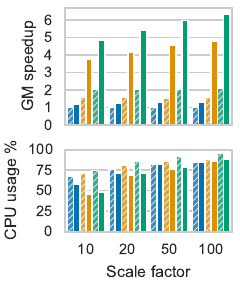}
        \caption{TPC-H}
    \end{subfigure}
    \caption{Geometric mean query speedup (relative to D1 without PIM, same scale factor) and average CPU usage for varying scale factor and denormalization level.}
    \label{fig:speedup-and-cpu-usage-vs-scale-factor}
\end{figure} 

\subsection{Speedup tends to increase as database size increases}

We now investigate the effect of database size on query speedup. Recall that the number of rows in each table is proportional to the scale factor, with the exception of the \texttt{part} table in SSB, which scales logarithmically. As shown in \Cref{tab:database-size-vs-scale-factor}, database size is roughly proportional to scale factor.

Varying the scale factor from 10 to 100, we evaluate Membrane's performance for denormalization levels D1-3, shown  in \Cref{fig:speedup-and-cpu-usage-vs-scale-factor}. We observe that query speedup tends to increase as database size increases. At scale factor 10 and denormalization level D3, the geometric mean query speedups are 4.4x and 5.0x for SSB and TPC-H. At scale factor 100, the speedups increase to 5.9x and 6.4x. Database size has little effect on query speedup without PIM.

To explain the effect of database size of query speedup, we measured average CPU usage for each configuration. We note that the CPU usage reported here excludes the period spent waiting for PIM filtering to complete. Results are shown in \Cref{fig:speedup-and-cpu-usage-vs-scale-factor}. At smaller scale factors, CPU usage with PIM is significantly lower than CPU usage without PIM. 

During query processing, database systems typically incur overheads for parsing, planning, optimization, and scheduling. Although DuckDB is extensively optimized, 
at small scale factors, the CPU has very little data left to process after PIM filtering, so these overheads play an outsized role.

\subsection{Speedup tends to increase as PIM selectivity decreases}

We now explore the impact of PIM selectivity on query speedup. We define PIM selectivity as the fraction of rows returned by PIM after filtering, or equivalently, the fraction of set bits in the bitmap. A given query's PIM selectivity may depend on the denormalization level. For example, TPC-H Q3 has a PIM selectivity of about 0.54 for D1 and 0.005 for D2. 

\begin{figure}
    \centering
    \includegraphics[width=2.6in]{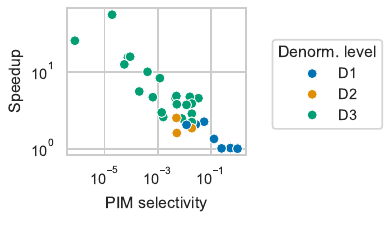}
    \caption{Query speedup for varying PIM selectivity and denormalization level (relative to D1 without PIM).}
    \label{fig:speedup-vs-pim-selectivity}
\end{figure}

In \Cref{fig:speedup-vs-pim-selectivity}, we show query speedup versus PIM selectivity for combined SSB and TPC-H. Each point in the plot is an individual query. We observe that query speedup tends to increase as PIM selectivity decreases. All queries with PIM selectivity less than $10^{-4}$ are at least 10x faster in Membrane. In contrast, among queries with PIM selectivity greater than 0.1, the maximum query speedup is 1.3x. Fortunately, analytical queries usually include filters with low selectivity, which can be accelerated in Membrane after partial denormalization. For D3, the minimum and maximum PIM selectivities are $7.6 \times 10^{-7}$ and $0.034$, respectively, and the minimum and maximum query speedups are 2.2x and 57x, respectively.

\section{Related Work} \label{sec:relwork}




Prior works in the database field such as BitWeaving \cite{bitweaving} exploited the ``intra-cycle''/bit-level parallelism of processors to accelerate the scan and 
filtering kernels. SIMD-scan \cite{simdScan} aimed to perform the same by utilizing on-chip vector processing units with SSE instructions. 

{\bf Processing In Storage Solutions.} 
With database machines \cite{databasemachines}, there were attempts  in the 1970s and 1980s to push query computation closer to where the data resided---at that time, spinning disks. 
However, these efforts were abandoned as the resulting custom storage package was expensive to manufacture and commodity microprocessors were seeing exponential growth in performance. 
However, with the slowing  of Moore's Law, there is a need to revisit ideas for specialization in today's context. Pinatubo \cite{pinatubo} and SmartSSD \cite{smartssd} are examples of other works that have proposed pushing query processing into the storage device. These designs, however, are limited by the storage I/O interface and suffer from higher latency and lower degrees of parallelism, and do not serve the needs of markets using in-memory databases. 

{\bf DRAM-Based PIM Designs.}
Several prior works such as Ambit \cite{ambit} and SIMDRAM \cite{simdram} propose a triple-row activation design to perform logical operations at the subarray-level that could be leveraged for processing OLAP queries, but these approaches require more significant changes to the DRAM, in particular support for multiple concurrent row activations per bit-level operation.
{JAFFAR \cite{jaffar} is a DIMM-level design that focuses on the filter operation by operating on the I/O buffer present on each DIMM. Although it gains by reducing data that travels over the memory bus, the amount of parallelism available in the I/O buffer is limited. This approach is similar to our rank-level approach. The Reconfigurable Vector Unit \cite{rvu} proposes to implement vector processing units at a vault-level in an HMC design. Polynesia \cite{polynesia} accelerates the analytical portion of HTAP database workloads using vault-level processing elements on HMC. Our approach would also extend to HMC or HBM, but in-memory databases benefit from the greater capacity scaling of conventional DIMMs. 
Most prior work also fails to evaluate end-to-end query processing pipelines. 
Membrane differs from most of these works in that it thoroughly explores the design space for conventional DIMM memory and cooperatively processes the entire query together with the host rather than offloading the entire query processing to PIM hardware.

{\bf Alternative Architectures.}
Prior works such as \cite{kevin_old_gpu_paper}  accelerated the filtering step on the GPU but omitted the data-retrieval portion and subsequent postprocessing, which we have shown will often consume a large portion of query processing time.
Ibex \cite{ibex} and \cite{fpga-sql-solution} implemented query processing on FPGAs. However, GPUs and FPGAs suffer from the limited scalability of onboard memory compared to the main memory addressable by the CPU. Papaphilippou and Luk~\cite{fpga-database-acceleration-survey} provides a comprehensive survey of works investigating acceleration of database systems using FPGAs and arrives at similar conclusions.

We implemented an optimized version of the filter kernel on an Alveo U280 FPGA to take  advantage of the onboard HBM memory to execute the filter microkernel described earlier in \Cref{sec:scan_microbenchmark}. We observe that Membrane outperforms this FPGA setup by at least 27.46x, including the cost of data transfers to and from the FPGA.


GPUs should be compatible with our Membrane-based cooperative processing approach. CPUs, discrete GPUs, and similar processing units can utilize Membrane bank-level PIM units for filter and transfer intermediate results efficiently back to the hosts. 
\section{Conclusions}


In-memory analytics can be accelerated by offloading the filter kernels to PIM processing units. In this work, we observe that denormalization methods make these workloads significantly more amenable to PIM filtering, albeit by incurring extra memory overheads. We evaluated different levels of denormalization that provide a tradeoff between increased memory consumption and improved performance. We thoroughly explored the DRAM design space to conclude that bank-level offers high performance with minimal area overhead and power usage. Membrane's bank-level PIM can outperform the CPU baselines by 5.9-6.3x and have a memory overhead of 9-17\%, depending on the different denormalization levels across both TPCH/SSB benchmarks. 
\section{Acknowledgements}
This work is funded in part by the National Science Foundation (NSF) under collaborative awards CCF-2312739, CCF-2312740, and CCF-2312741, as well as  PRISM and ACE, two of seven centers in JUMP 2.0, a Semiconductor Research Corporation (SRC) program, sponsored by DARPA. 


\bibliographystyle{IEEEtranS}
\bibliography{refs.bib}

\end{document}